\documentclass[twocolumn,prl,showpacs,nofootinbib]{revtex4}

\usepackage{graphicx}
\usepackage{dcolumn}
\usepackage{bm}

\newcommand{\beq}{\begin{equation}}
\newcommand{\eeq}{\end{equation}}
\newcommand{\beqa}{\begin{eqnarray}}
\newcommand{\eeqa}{\end{eqnarray}}

\begin{document}

\title{Naked-eye optical flash from GRB 080319B: Tracing
the decaying neutrons in the outflow}

\author{Yi-Zhong Fan}
\email{yizhong@nbi.dk}%
\affiliation{%
Niels Bohr International Academy, Niels Bohr Institute,
Blegdamsvej 17, DK-2100 Copenhagen, Denmark\\
Purple Mountain Observatory, Chinese Academy of Sciences, Nanjing
210008, China}%
\author{Bing Zhang}
\affiliation{%
Department of Physics and Astronomy,
   University of Nevada, Las Vegas,
NV 89154, USA.}%
\author{Da-Ming Wei}
\affiliation{%
Purple Mountain Observatory, Chinese Academy of Sciences, Nanjing
210008, China.}%

\date{\today}

\begin{abstract}
For an unsteady baryonic gamma-ray burst (GRB) outflow, the fast and
slow proton shells collide with each other and produce energetic
soft gamma-ray emission. If the outflow has a significant neutron
component, the ultra-relativistic neutrons initially expand freely
until decaying at a larger radius. The late time proton shells
ejected from the GRB central engine, after powering the regular
internal shocks, will sweep these $\beta-$decay products and give
rise to very bright UV/optical emission. The naked-eye optical flash
from GRB 080319B, an energetic explosion in the distant universe,
can be well explained in this way.
\end{abstract}


\maketitle

There were four gamma-ray bursts (GRBs) detected on 19 March, 2008.
Among them, GRB 080319B was most noticeable due to its huge
isotropic energy and its extremely bright prompt optical emission
that could have been seen with naked eye \cite{Racu08}. For a
redshift $z = 0.937$, the detected peak optical emission had a
visual magnitude $\sim 5.3$ that corresponds to an optical
luminosity $\geq 5\times 10^{50}~{\rm erg~s^{-1}}$. The simultaneous
soft $\gamma-$ray emission had a luminosity $L_\gamma \sim 4\times
10^{52}~{\rm erg~s^{-1}}$. The optical flux is significantly above
the spectral extrapolation of the GRB emission into the optical band
\cite{Racu08}, suggesting that the optical and the gamma-ray
emissions originate from different emission sites or belong to
different spectral components at a same emission site. The
lightcurves in the two bands show some similarities, but do not
trace each other exactly \cite{Racu08}. This suggests that the
emissions from the two bands are somewhat related to each other.

For the energetic outflow of GRB 080319B, if not strongly
magnetized, a significant neutron component is unavoidable
\cite{DKK99,FPA00,Pruet03,Bel03}. The average Lorentz factor of the
outflow before getting decelerated by a stellar wind medium is very
high. A lower limit can be set by the Lorentz factor of the forward
shock at $\sim 70$ s, when the X-ray afterglow began to decline
normally, i.e., $\Gamma>500 E_{\rm
k,55}^{1/4}A_{*,-2}^{-1/4}(t/70{\rm s})^{-1/4}[(1+z)/2]^{1/4}$,
where $E_{\rm k}\sim 10^{55}~{\rm erg}$ is the isotropic-equivalent
energy of the outflow and $A_* \sim 0.01$ is the stellar wind
parameter of the progenitor \cite{Racu08}. The current data then
suggests an initial Lorentz factor of the outflow $\Gamma \sim
10^3$. For an unsteady baryonic outflow, the GRB is powered by the
interaction of proton shells with variable Lorentz factors, i.e.,
internal shocks \cite{PX94,RM94,Piran99,Mesz02,ZM04}. In order to
convert a significant fraction of the initial kinetic energy into
internal energy and then to $\gamma-$ray radiation, the difference
in Lorentz factors between the shells should be substantial (i.e.,
the Lorentz factors $\eta_{\rm f}\gg \eta_{\rm s}$, hereafter the
subscript ``f'' and ``s'' denote ``fast'' and ``slow'',
respectively) and their masses should satisfy $M_{\rm f}=f M_{\rm
s}$ with an $f> \eta_{\rm s}/\eta_{\rm f}$. The mergered new proton
shell moves with a Lorentz factor $\Gamma_{\rm m}\approx \sqrt{f
\eta_{\rm f}\eta_{\rm s}}$ \cite{Piran99}. For $\eta_{\rm f}\sim
{\rm a ~few~} \times 10^4$ and $f\sim 0.1$, $\eta_{\rm s}\sim {\rm
a~few~} \times 100$ is needed to get a $\Gamma_{\rm m} (\sim \Gamma)
\sim 10^3$.

The dynamics of a neutron-rich outflow is governed by the
dimensionless entropy $\eta=L/(\dot{M}c^2)$ at $r_0$, where
$\dot{M}$ is the mass loading rate, $L$ is the isotropic luminosity
of the ejecta, and $r_0$ is the radius of the central engine.
Whether or not the proton and the neutron components decouple from
each other depends on whether $\eta$ is above or below the critical
value $\eta_{\rm cr}\simeq 10^3 L_{54}
^{1/4}r_{0,7}^{-1/4}[(1+\xi)/2]^{-1/4}$, where $\xi$ is the ratio of
the number density of neutrons to protons \cite{BM00}. The
convention $Q_{ x}=Q/10^{ x}$ is adopted in this work in cgs units.
For $\eta<\eta_{\rm cr}$, the neutron and proton components (denoted
by the subscripts $n$ and $p$, respectively) are still coupled with
each other by nuclear elastic scattering at a radius $\sim \eta
r_0$, i.e., at the end of the fireball acceleration. Thus for slow
shells one usually has $\eta_{n,s}=\eta_{p,s}$. For $\eta>\eta_{\rm
cr}$, the $n$ and $p$ components decouple at a radius $\sim
\eta_{n,f} r_0$ where the nuclear elastic scattering becomes too
weak to accelerate the neutrons, where $\eta_{n,f}\sim 10^3
L_{54}^{1/4} r_{0,7}^{-1/4}[(1+\xi)/2]^{-1/4}(\eta_{p,f}/\eta_{\rm
cr})^{-1/3}$ \cite{BM00}. For the observed information of GRB
080319B, we take $\eta_{p,s}=\eta_{n,s} (\sim \eta_{n,f}) \sim {\rm
a ~few~hundred}$, $\eta_{p,f} \sim {\rm a~few}\times 10^4$, $M_{\rm
p,s} \sim 10M_{\rm p,f}$, and $\xi\sim 1$. Hereafter $\Gamma_{\rm
n}$ denotes the generic Lorentz factor of the neutron shell
regardless of whether it is relatively ``slow'' or ``fast''. Below
we show that with these parameters, both the soft $\gamma-$ray and
optical emission can be interpreted.

Firstly we discuss the regular internal shocks powered by the
collisions of the fast and slow proton shells at a radius $R_{\rm
int}\sim 2\Gamma_{\rm m}^2 c\delta t/(1+z) \sim 3\times 10^{14}~{\rm
cm}~\Gamma_{\rm m,3}^2 \delta t_{-2}[2/(1+z)]$ cm, where $\delta t$
is the typical variability timescale of the prompt $\gamma-$ray
light curve. The electrons are accelerated by the internal reverse
shock to a typical Lorentz factor \cite{Piran99} $\sim 5\times
10^3(\epsilon_e/0.3)({\eta_{p,f}}/2\times 10^4)\Gamma_{\rm
m,3}^{-1}$ and the magnetic field generated in the shocks can be
estimated as $\sim 4\times 10^{3}~{\rm
Gauss}~(3\epsilon_B/\epsilon_e)^{1/2}L_{\gamma,52.6}^{1/2}R_{\rm
int,14.5}^{-1}\Gamma_{\rm m,3}^{-1}$, where $\epsilon_e$ and
$\epsilon_B$ are the fractions of the shock energy distributed to
electrons and magnetic fields, respectively. The typical synchrotron
emission frequency of these electrons in the internal reverse shock
is $\sim 1.4\times 10^{20}$ Hz, matching the observation. The
internal forward shock can only accelerate electrons to a typical
Lorentz factor $\sim 300$ and the emission is in the soft X$-$ray
band.

The neutrons have negligible interaction with the protons before
decaying into protons, electrons and electron neutrinos. The
$\beta-$decay radius reads
\begin{equation}
R_{\beta} \approx 1.1\times10^{16}~{\rm cm}~\Gamma_{\rm n,2.6}.
\end{equation}
A pair of fast/slow proton shells ejected at late times would merge
into a proton shell with a Lorentz factor $\sim \Gamma_{\rm m}$ in
the inner internal shocks, and then catch up with the decay trail of
the neutron shells ejected earlier at a radius $R_{\rm cat} \approx
2\Gamma_{\rm n}^2 c\delta T/(1+z)$, where $\delta T$ is the ejection
time-lag between the earlier neutron shell and the later proton
shell. As long as $R_{\rm cat} \geq R_\beta$, which requires $\delta
T \geq 1.1(1+z){\Gamma_{\rm n,2.6}^{-1}}$ sec, there will be a
substantial amount of $\beta-$decay products that would be swept
orderly by the later proton shell at a radius of $\sim 2 R_\beta$.
These ``secondary internal shocks" are unable to give rise to
energetic X$-$ray and $\gamma-$ray emissions for the following
reasons. (1) They are generated at a radius much larger than $R_{\rm
int}$, so that the magnetic fields are much weaker than those in the
regular internal shocks. (2) They have much lower efficiency than
the inner ones at $R_{\rm int}$ because of the smaller Lorentz
contrast between the merged proton shell and the neutron shell. For
the same reason, the total energy converted into internal energy and
then into radiation is lowered by a factor of $> 10$ than that of
the regular internal shocks.

\begin{table}
\caption{Physical parameters that reproduce the prompt emission of
GRB 080319B.}
\begin{tabular}{lll}
\hline
Quantity & ~~slow shell & ~~fast shell  \\
\hline
Lorentz factor of protons & ~~$\sim 400$ & ~~$\sim 2\times 10^{4}$ \\
Lorentz factor of neutrons & ~~$\sim 400$ & ~~$\sim 500$\\
isotropic luminosity (erg/s) & $\sim 2\times 10^{53}$ & ~~$\sim 10^{54}$\\
mass ratio of neutrons to protons & ~~$\sim 1$ & ~~$\sim 1$ \\
entropy per baryon ($s/k$) & ~~$\sim 1.2\times 10^5$ & ~~$\sim 4\times 10^6$\\
\hline
\end{tabular}
\label{Tab:080319B}
\end{table}

We first look at the interaction between a proton shell formed in
the inner internal shocks and the decay trail of a series of
identical neutron shells. The number density of the decay products
being swept by the proton shell reads \cite{fw04} $n\approx
{\Gamma_{\rm n}M_{n}\over 2\pi R^2m_{\rm n}R_{ \beta}}$, where
$m_{\rm n}$ is the neutron rest mass, and $M_{n}\propto \exp(-R/R_{
\beta})$ is the rest mass of the neutron shell that undergos
$\beta-$decay. The minimum Lorentz factor of the shocked electrons
can be estimated as \cite{Piran99} $\gamma_{e, m} \approx 55
({\epsilon_e\over 0.3}){3(p-2)\over (p-1)}({\gamma_{\rm rel}-1\over
0.3})$, where $p$ is the power-law index of the accelerated
electrons, and $\gamma_{\rm rel}\approx
(\Gamma/\Gamma_n+\Gamma_n/\Gamma)/2$ is the Lorentz factor of the
trailing fast-moving proton shell with a Lorentz factor $\Gamma$
relative to the neutron shell. The magnetic field strength $B'$ can
be estimated as \cite{spn98}
\begin{eqnarray}
B'&\sim & 60~{\rm Gauss}~\epsilon_{\rm B,-1}^{1/2}(\gamma_{\rm
rel}^2-1)^{1/2}N_{n,51.5}^{1/2}\nonumber\\
&& R_{16}^{-1}R_{\beta,16}^{-1/2} \Gamma_{\rm
n,2.6}^{1/2}\exp(-R/2R_\beta),
\end{eqnarray}
where the neutron number $N_{n}$ of one shell is estimated by
$N_{n}\sim \xi L \delta t/[(1+z)\Gamma_{\rm m} m_{\rm n}c^2]\sim
3.3\times 10^{51}~ \xi L_{54}\delta t_{-2}\Gamma_{\rm m,3}^{-1}$.
Since this secondary internal shock region is permeated by gamma-ray
photons produced from the ``inner'' internal shocks from the
late-time ejected proton shells, the electrons in the secondary
internal shock region suffer Compton cooling by these prompt
$\gamma-$rays \cite{fzw05}. The corresponding cooling Lorentz factor
reads
\begin{equation}
\gamma_{e,c} \sim 180~\Gamma_{3}^3R_{16}L_{\gamma,52.7}^{-1}.
\label{eq:gamma_c}
\end{equation}

So, the synchrotron radiation energy of the shocked electrons peaks
at a frequency
\begin{equation}
\nu_{\rm p} \sim 1.4\times 10^{15}~{\rm Hz}~ \Gamma_3 B'_{2}
{\min\{\gamma_{e,c}^2,~\gamma_{e,m}^2\}\over 10^4}.
\end{equation}

The detected maximum specific spectral flux can be estimated as
\cite{spn98}
\begin{equation}
F_{\nu_{\rm max}} \sim 50~{\rm Jy}~N_{e,53}\Gamma_3 B'_2,
\end{equation}
where $N_e \sim N_n \delta t_\beta/\delta t\simeq 2\times
10^{53}~\xi L_{54}\Gamma_{\rm m,3}^{-1}\Gamma_3^{-2}R_{\beta,16}$,
and $\delta t_\beta \approx (1+z)R_\beta/\Gamma^2 c \sim 0.6~{\rm
s}~({1+z \over 2})R_{\beta,16}\Gamma_3^{-2}$. This flux is bright
enough to well exceed the spectral extrapolation of the gamma-ray
emission and to interpret the observed naked-eye optical flash.
Notice that we have introduced the number of proton shells ($\sim
\delta t_\beta/\delta t$) ejected during the time span of $\delta
t_\beta$ to account for the total emission output \cite{fw04}. Since
$F_{\nu_{\rm max}} \propto~N_{e}B' \propto \xi^{3/2}$, $\xi \sim 1$
is highly needed to reproduce the prompt optical emission with a
flux $\sim 20$ Jy.

Following the standard approach \cite{RL79}, we estimate the
synchrotron-self absorption frequency
\begin{equation}
\nu_a \sim 2\times10^{14}~{\rm Hz}~{100 \over
\min\{\gamma_{e,c},\gamma_{e,m}\}}N_{n,51.5}^{3/5}R_{16}^{-6/5}{B'}_{2}^{2/5}\Gamma_3.
\end{equation}
We can see that for the standard parameters, in particular the large
neutron decay radius $R_{16} \sim 1$ and the small $B'_2<1$, one has
$\nu_a < \nu_{\rm p}$. This is the main reason that the optical
flash of this burst is so bright. The lack of bright optical flash
in most other bursts may be attributed to their smaller Lorentz
factors, which give a smaller neutron decay radius, stronger
magnetic fields, and hence, a higher self-absorption frequency than
optical. In addition, smaller $\Gamma$ and $R_\beta$ would suppress
the optical emission also by reducing $\gamma_{e,c}$, as shown in
eq.(\ref{eq:gamma_c}).

Till now we have shown that with reasonable parameters (see
Tab.\ref{Tab:080319B} for a summary), the neutron-rich internal
shocks can power both the energetic soft $\gamma-$ray flare and the
extremely bright optical flash of GRB 080319B. The thermodynamic
entropy per baryon of the initial ejecta can be estimated as $s/k
\sim \eta m_{\rm n}c^2/(kT)$ \cite{Pruet03,Pruet02}, where $k$ is
the Boltzman's constant and $T\sim 5~{\rm
MeV}~r_{0,7}^{-1/2}L_{54}^{1/4}$ is the temperature of the initial
fireball. For the fast and slow shells, we have $s/k \sim (4\times
10^6,~ 1.2\times 10^5)$, respectively.

Since the neutron shells are originally coupled to the early proton
shells, they carry the essential variability information of the
early proton outflows. When the decay products are swept by a later
injected proton shell, the resultant optical lightcurve generally
follow the variability pattern of the gamma-ray lightcurve, yet does
not strictly trace the gamma-ray lightcurve. The time delay between
the optical and gamma-ray peaks can be estimated as $\sim
2.2[(1+z)/2]\Gamma_{\rm n,2.6}^{-1}$ sec, matching the time lag
$\sim 3$ sec found in the correlation analysis of prompt
$\gamma-$ray and optical lightcurves \cite{KP08}. Being at a larger
emission radius $\sim R_\beta$, the optical variability is also
smoothed by the geometric effect in a timescale $\sim \delta t_\beta
\gg \delta t$, generally consistent with the fact that the optical
lightcurve is much smoother than the gamma-ray lightcurve
\cite{Convino08}. The contamination from a bright external reverse
shock optical emission component \cite{Racu08} introduces a bright
background and also contributes to the smoothing.

In the synchrotron self-Compton (SSC) internal shock model
\cite{Racu08,KP08}, one requires a large emission radius ($\sim
10^{16}~{\rm cm}$) to avoid self-absorption, and also requires a
large $\epsilon_e/\epsilon_B~(> 10^3)$ to interpret the huge energy
contrast between the optical and the $\gamma-$ray emission. In that
scenario, one predicts a very bright prompt GeV emission component
with luminosity $\geq 4\times 10^{53}~{\rm erg~s^{-1}}$, which may
suffer an energy crisis \cite{PSZ08}. In our scenario, the SSC
component of the prompt GeV emission is most likely weaker because
our standard model parameters demand $\epsilon_e \sim \epsilon_B$.
GeV emission can also arise from the prompt $\gamma-$ray cooling in
the shocked region at $\sim R_\beta$. However, as mentioned before,
the total energy of the shocks at $\sim R_\beta$ is much smaller
than that of the ``inner" internal shocks. So the high energy
emission would not be enhanced significantly. Prompt UV/optical
photons will cool the forward shock electrons effectively and give
rise to GeV-TeV emission with Luminosity $\sim 10^{52}~{\rm
erg~s^{-1}}$. However, such a component, appearing as a plateau,
should last two more times longer than the prompt optical emission
and thus can be easily distinguished \cite{FP08}.

\begin{table*}
\caption{General features of the neutron-rich internal shock model
for GRB 080319B.}
\begin{tabular}{lll}
\hline
Quantity & ~~regular internal shocks & ~~secondary internal shocks  \\
\hline
slow material & slow protons from central engine& ~~$\beta-$decay products of neutrons\\
shock radius & ~~$\sim 10^{14}-10^{15}$ cm &  ~~$\sim R_\beta \sim 10^{16}$ cm \\
shock strength & ~~ultra-relativistic & ~~sub-relativistic  \\
typical emission  & ~~$\gamma-$rays & ~~ultraviolet/optical photons  \\
emission flux & ~~$\sim 10^{-5}$ erg/s/cm$^2$ & ~~$\sim 50$ Jy (in optical band) \\
variability timescale & ~~$\delta t \sim 0.01-0.1$ sec & ~~$\delta t_{\beta} \sim 1$ sec \\
\hline
\end{tabular}
\label{Tab:080319B2}
\end{table*}

In summary, we show in this work that an outflow, containing
comparable amounts of protons and neutrons (i.e., $\xi \sim 1$) and
having an averaged Lorentz factor $\sim 10^{3}$ and a huge isotropic
luminosity $L\sim 10^{54}~{\rm erg~s^{-1}}$, can power an energetic
soft $\gamma-$ray burst and a naked-eye optical flash
in GRB 080319B (see Tab.\ref{Tab:080319B2} for a summary). In our
scenario, the prompt optical light curve will be much smoother than
that of the prompt $\gamma-$ray emission and the peaks in optical
band will lag behind the $\gamma-$ray peaks by a few seconds, both
are well consistent with the data \cite{KP08,Convino08}
and are hard to be interpreted in other models that invoke a similar
emission radius for both prompt gamma-ray and
optical emissions \cite{Racu08,KP08,Yu08}.\\

\acknowledgments YZF is supported in part by Danish National
Research Foundation, Chinese Academy of Sciences and National Basic
Research Program of China (grant 2009CB824800). BZ is supported by
NASA NNG05GB67G and NNX08AE57A. DMW is supported by the National
Natural Science Foundation of China (Grants 10621303, 10673034) and
National Basic Research Program of China (973 Program 2007CB815404).

\end{document}